\documentclass[a4paper]{jpconf}
\usepackage{graphicx}
\begin{document}
\title{Photorecombination of berylliumlike Ti$^{18+}$: Hyperfine quenching of dielectronic resonances}

 \author{S Schippers$^1$, E W Schmidt$^1$, D Bernhardt$^1$, D Yu$^{1,}$\footnote[3]{Permanent address: Institute of Modern Physics,
           Chinese Academy of Sciences, Lanzhou 730000, P. R. China}, A
 M\"{u}ller$^1$,\\ M Lestinsky$^2$, D A Orlov$^2$, M Grieser$^2$, R Repnow$^2$, A Wolf$^{\,2}$}

 \address{$^1$Institut f\"{u}r Atom- und Molek\"{u}lphysik, Justus-Liebig-Universit\"{a}t, 35392
  Giessen, Germany\\
  $^2$Max-Planck-Institut f\"{u}r Kernphysik, 69117 Heidelberg, Germany}

\ead{Stefan.E.Schippers@iamp.physik.uni-giessen.de}

\begin{abstract}
The photorecombination spectrum of $^{48}$Ti$^{18+}$ was measured
employing the merged electron-ion beams technique at a heavy-ion
storage ring. The experimental electron-ion collision energy range
$0-80$~eV comprises all dielectronic recombination (DR) resonances
associated with $2s\to2p$ ($\Delta N =0$) core excitations as well
as trielectronic recombination (TR) resonances that involve
$2s^2\to2p^2$ core double-excitations. At low collision energies
DR resonances are observed that are associated with the excitation
of metastable $2s\,2p\,^3P_0$ primary ions with nearly infinite
lifetime for the isotope $^{48}$Ti with zero nuclear spin. For the
isotope $^{47}$Ti with nonzero nuclear spin hyperfine quenching of
these resonances occurs. The procedure for obtaining the
associated time constant from a recombination measurement is
outlined.
\end{abstract}

\section{Introduction}

The $ns\,np\;^3P_0$ state in divalent atoms and ions is the first excited
state above the $ns^2\;^1S_0$ ground state. Its decay by a one-photon
$J=0 \to J=0$ transition is strictly forbidden and the lifetime of the
$^3P_0$ state is nearly infinite provided the ion has zero nuclear spin
$(I=0)$. If $I>0$ the hyperfine interaction mixes levels with different
$J$ and, consequently, the $^3P_0$ level acquires a finite lifetime. This
hyperfine quenching has been treated theoretically for Be-like, Mg-like
and Zn-like ions \cite{Marques1993,Marques1993a,Brage1998a,Liu2006a}.
Theoretical work on neutral alkaline-earth-metal atoms
\cite{Porsev2004a,Santra2004a} was motivated by the idea to use the
hyperfine induced $^3P_0 \to {^1S_0}$ transition in ultraprecise atomic
clocks. In another application the $^{13}$C$^{2+}$($^3P_0 \to {^1S_0}$)
fluorescence from planetary nebulae was measured to infer the
$^{13}$C/$^{12}$C abundance ratio and, thereby, to gain insight into
stellar nucleosynthesis \cite{Rubin2004a}.

So far, the only experimental values for $ns\,np\;^3P_0 \to
ns^2\;^1S_0$ hyperfine induced (HFI) transition rates were
obtained for In$^+$ ($n$=5) \cite{Becker2001a} stored in a radio
frequency trap and for Be-like N$^{3+}$ from astrophysical
observations \cite{Brage2002a}. Although the latter result had a
rather large uncertainty of $\pm33\%$, it allows one to
discriminate between the two theoretical values
\cite{Marques1993,Brage1998a} that are available for this ion and
that differ by almost a factor of 4. In view of these theoretical
uncertainties it is evident that accurate experimental benchmarks
are highly desirable, especially for few electron systems such as
Be-like ions where correlation effects are particularly strong.

In the present work the hyperfine quenching of the $2s\,2p\;^3P_0$ state
of Be-like $^{47}$Ti$^{18+}$ was observed in dielectronic recombination
(DR) experiments carried out at the heavy-ion storage-ring TSR of the
Max-Planck-Institute for Nuclear Physics in Heidelberg, Germany. The
procedure for obtaining the associated decay constant from electron-ion
recombination measurements is outlined. The result will be published
elsewhere \cite{Schippers2006a}. Here, the Ti$^{18+}$ recombination
spectrum is presented and its isotope-dependence is discussed.

\section{Experiment}

Mass selected $^{47,48}$Ti$^{18+}$ ion beams (natural abundances
7.2\% and 73.7\%, respectively) were provided by a tandem
accelerator, followed by a radio frequency linear accelerator, at
energies close to 240 MeV, using a fixed magnetic setting for the
beam line and the storage ring. In one straight section of the
storage ring the ion beam was continuously phase-space cooled
using the velocity-matched electron beam of the TSR electron
cooler. In a second straight section, the ion beam was merged with
the collinear electron beam of the high-resolution electron target
\cite{Sprenger2004a}, run at variable acceleration voltage in
order to set the required collision energy in the co-moving
reference frame of the ions. The target electron beam temperatures
were $kT_\| \approx 40$~$\mu$eV and $kT_\perp\approx 4$~meV.
Ti$^{17+}$ ions formed by electron-ion recombination in the
electron target or by charge transfer in collisions with residual
gas molecules were deflected out of the closed orbit of the
circulating Ti$^{18+}$ ion beam in the first dipole magnet
downstream of the electron target and were directed onto a
scintillation detector operated in single-particle counting mode
with nearly 100\% detection efficiency.

Recombination spectra of the Ti$^{18+}$ ions as a function of the
relative electron-ion energy were taken by varying the cathode
voltage of the electron target appropriately.  The procedure for
electron-ion measurements at the TSR storage ring has been
described in more detail in, e.\,g., Ref.\ \cite{Schippers2000b}
(and references therein).  For the present spectral measurements,
a constant current of cooled, circulating Ti$^{18+}$ ions was
maintained.  Currents of a fraction of a $\mu$A were injected at a
rate of $\approx$1 s$^{-1}$ in order to obtain stationary stored
currents of $\approx$40~$\mu$A for $^{48}$Ti$^{18+}$ and
$\approx$6~$\mu$A for $^{47}$Ti$^{18+}$, respectively; this
largely reflects the difference in the natural isotope abundances.

\section{Results and discussion}

\begin{figure}[t]
\includegraphics[width=0.775\textwidth]{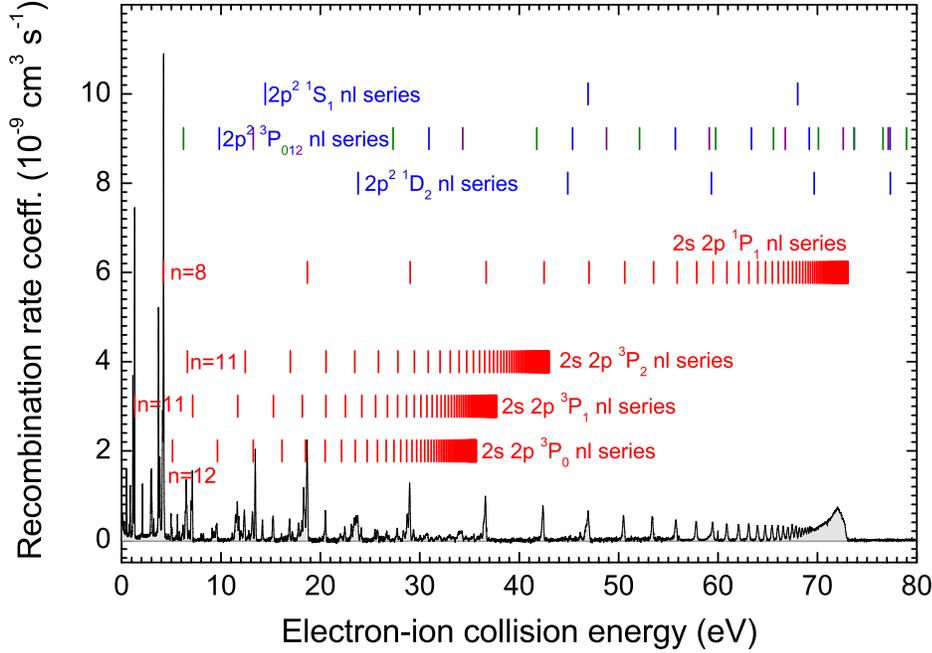}
\begin{minipage}[b]{0.215\textwidth}\caption{\label{fig:overview}Measured recombination spectrum of
berylliumlike $^{48}$Ti$^{18+}$. The vertical lines indicate DR
resonance positions $E_n$ for excitation from the $2s^2\;^1S_0$
ground state that were estimated using the Rydberg formula $E_n =
E_\infty - 13.606\mathrm{~eV}\times(18/n)^2$ with the series limits
$E_\infty$ taken from the NIST atomic spectra data base
\cite{Ralchenko2005a}.}
\end{minipage}
\end{figure}

An overview over the measured $^{48}$Ti$^{18+}$ recombination spectrum is
presented in figure \ref{fig:overview}. The energy range 0--80~eV
comprises all resonances associated with $2s\to2p$ ($\Delta N=0$) core
excitations. The spectrum is similar to the previously measured ones of
the isoelectronic ions Cl$^{13+}$ \cite{Schnell2003b} and Fe$^{22+}$
\cite{Savin2006a}. At higher energies it is dominated by the regular
Rydberg series of $2s^2\;^1S_0\to (2s\,2p\;^1P_1)\,n\ell$ DR resonances
converging to the $2s\,2p\;^1P_1$ series limit at 73.11~eV
\cite{Ralchenko2005a}. The $2s\,2p\;^3P_J$ series limits are barely
visible. They are expected to occur at 35.73, 37.77, and 43.05~eV for
$J=0$, 1, and 2, respectively. At lower energies the spectrum is more
irregular due to the larger fine-structure splitting of the lower-$n$
resonances. Some resonance features, e.\,g., at 23.6 and 34.0~eV can be
related to $2s^2\to2p^2$ core double-excitations. A thorough discussion
of these trielectronic recombination (TR) resonances can be found
elsewhere \cite{Schnell2003b}.

In single-pass merged-beams electron-ion recombination experiments
with lower charged Be-like C$^{2+}$, O$^{4+}$, and F$^{5+}$ ions,
strong Rydberg series of DR resonances were observed associated
with $2s\,2p\;^3P \to 2s\,2p\;^1P$ and $2s\,2p\;^3P \to 2p^2\;^3P$
core excitations of metastable $2s\,2p\;^3P$ primary ions
\cite{Badnell1991a}. In these experiments the metastable $^3P$
fraction in the primary ion beams amounted up to $\approx$70\%. In
the present $^{48}$Ti$^{18+}$ recombination spectrum analog strong
resonance features are absent. The $J=1$ and $J=2$ levels of the
Ti$^{18+}$($2s\,2p\;^3P)$ term have theoretical lifetimes of 74~ns
and 1~ms, respectively \cite{Bhatia1980a}, much shorter than the
average storage lifetime ($\approx$50 s). Thus, the major fraction
of the initially present metastable ions decayed to the ground
state very fast on the time scale of the experiment, and the
remaining (infinitely) long-lived $^3P_0$ metastable fraction
amounted to only $\approx$5\% (see below). In the single-pass
experiments, however, typical flight times were of the order of
several $\mu$s, i.\,e., much shorter than the lifetimes of the
$^3P_1$ and $^3P_2$ states. For lower charged ions these lifetimes
are even larger than the values given above, e.\,g.\, 0.16~ms
(theoretical value \cite{Marques1993}) for the
F$^{5+}$($2s\,2p\;^3P_1$) state, so that even the fastest decaying
$^3P_1$ component of the $^3P$ term was contained in the primary
beams of the single-pass experiments.

\begin{figure}[b]
\includegraphics[width=0.7\textwidth]{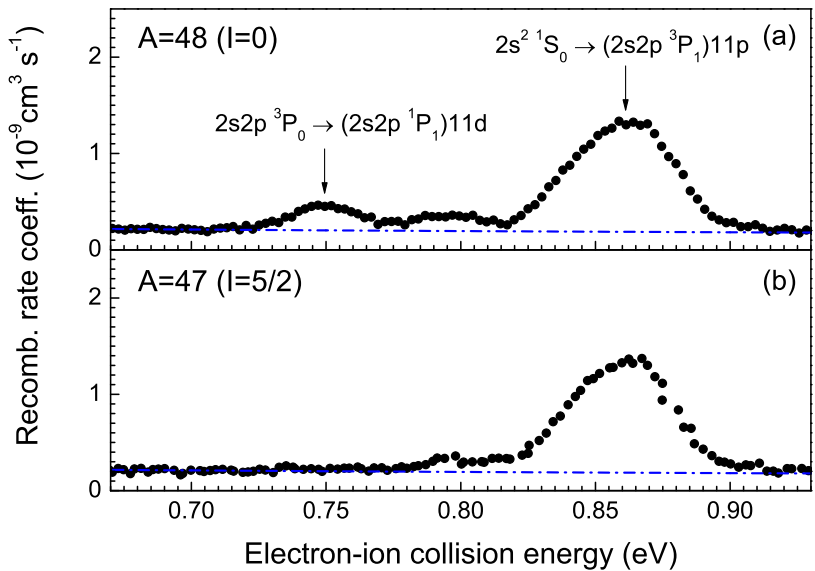}
\begin{minipage}[b]{0.29\textwidth}\caption{\label{fig:quenching} Detailed view of DR resonance structures
at low energies: (a) $^{48}$Ti$^{18+}$  and (b) $^{47}$Ti$^{18+}$.
The dash-dotted line is the theoretical rate coefficient for
nonresonant radiative recombination. Resonances associated with DR
of metastable $^3P_0$ ions are not observed when $^{47}$Ti$^{18+}$
ions with nonzero nuclear spin $I$ are continuously injected into
the storage ring.}
\end{minipage}
\end{figure}

In the present experiment DR resonances associated with the
excitation of the metastable $^3P_0$ fraction of the primary ion
beam are only observed at low electron-ion collision energies.
Figure \ref{fig:quenching}(a) shows a region of the recombination
spectrum where resonances of metastable
$^{48}$Ti$^{18+}$($2s2p\;^3P_0$) ions occur close to a strong
resonance from ground-state Ti$^{18+}$ ions. Theoretical
calculations using the \textsc{autostructure} code
\cite{Badnell1986} were performed \cite{Schippers2006a} to assign
the weaker structures to the metastables. Additionally, the
calculations indicate an average population of $\approx$5\% for
the metastable Ti$^{18+}$($2s2p\;^3P_0$) ions in the stored
$^{48}$Ti$^{18+}$ beam.

Using a $^{47}$Ti$^{18+}$ beam  the resonances assigned to metastable
Ti$^{18+}$($2s2p\;^3P_0$) ions essentially disappear [Fig.\
\ref{fig:quenching}(b)]. Their average population is strongly reduced
through the  $2s\,2p\;^3P_0\to 2s^2\;^1S_0$ HFI radiative decay. The
isotope-dependent occurrence of resonances in a DR spectrum was also
observed in earlier TSR experiments, using the heavier divalent ion
Pt$^{48+}$ (Zn-like) \cite{Schippers2005b}.  Here, resonances due to
$4s\,4p\;^3P_0$ were seen for $^{194}$Pt$^{48+}$ and disappeared for
$^{195}$Pt$^{48+}$. A measurement of the optical decay lifetime,
predicted to be 0.2971~ms \cite{Marques2006a}, was not attempted as, in
contrast to the present case, it should be much shorter than the time
needed for electron cooling of the ion beam. In the present case of
$^{47}$Ti$^{18+}$ the theoretical value for the HFI decay rate of the
$^3P_0$ state is $\tau_\mathrm{HFI} = 2.8$~s \cite{Marques1993} so that a
sizeable $^3P_0$ fraction can be expected in the ion beam even after
cooling for some 100 ms.

For the experimental determination of $\tau_\mathrm{HFI}$ the decay of
the $^{47}$Ti$^{18+}$($^3P_0$) beam component was monitored as a function
of storage time. To this end the relative electron-ion energy in the
electron target was set fixed to 0.75~eV where the $2s\,2p\;^3P_0\to
(2s\,2p\;^1P_1)\,11d$ DR resonance associated with excitation of the
$^3P_0$ state occurs (figure \ref{fig:quenching}). After injection of a
single Ti$^{18+}$ ion pulse into the storage ring the recombination rate
was recorded for up to 200~s. Prior to the injection of the next pulse
the remaining ions were kicked out of the ring. This scheme was repeated
for a sufficient number of times to reduce statistical uncertainties to a
level as low as achievable within one week of beam time. This method had
been applied previously for measuring the slow radiative decay rates of
$1s\,2s\;^3S$ states in the He-like ions (B$^{3+}$, C$^{4+}$, N$^{5+}$
\cite{Schmidt1994} and Li$^{+}$ \cite{Saghiri1999}). The experimental
result for $\tau_\mathrm{HFI}$ of the $^{47}$Ti$^{18+}$($2s\,2p\;^3P_0$)
state will be presented and discussed in a forthcoming publication
\cite{Schippers2006a}.

\ack

The authors thank M. Schnell for a helpful discussion and gratefully
acknowledge the excellent support by the MPI-K accelerator and TSR crews.
This work was supported in part by the German federal research-funding
agency DFG under contract no.\ Schi~378/5.

\medskip

%\bibliographystyle{iop}
%\bibliography{/tex/schippers}

%\end{document}

\end{document}